\documentclass{PoS}
\usepackage{psfig}

\title{On the continuation of the dual amplitude with Mandelstam analyticity off
mass shell}

\ShortTitle{On the continuation of the dual amplitude with Mandelstam analyticity...}

\author{\speaker{Volodymyr Magas}\\%
Departament d'Estructura i Constituents de la Materia,
University of Barcelona,\\
Av. Diagonal 647, 08028 Barcelona, Spain\\
        E-mail: \email{vladimir@ecm.ub.es}}

\abstract{A model for the $Q^2$-dependent modified dual amplitude 
with Mandelstam analyticity (M-DAMA) is proposed. M-DAMA preserves all the attractive
properties of DAMA, such as its pole structure and Regge asymptotics, and
leads to a generalized 
dual amplitude $A(s,t,Q^2)$. This amplitude 
can be checked in the known kinematical limits, i.e. it should 
reduce to onshell
hadronic scattering amplitude for $Q^2=0$ and can be related to the nuclear structure function for $t=0$. 
In such a way we complete a
unified "two-dimensionally dual" picture of strong interaction. 
 By comparing the structure function $F_2$, resulting from
 M-DAMA, with  phenomenological parameterizations,
  we fix the $Q^2$-dependence in M-DAMA. 
 In all studied regions, i.e. large and low $x$
limits as well as for the resonance region, the results of M-DAMA are in
qualitative agreement with the experiment. The new feature of the M-DAMA is the possible existence, according to vector meson dominance, of $Q^2$ poles in negative $Q^2$ region,
accessible  in  $e^+ e^-$ annihilation.}

\FullConference{Diffraction 06, International Workshop on Diffraction in
High-Energy Physics\\
                 September 5-10, 2006\\
                 Adamantas, Milos island, Greece}

\begin{document}

\newcommand{\dlt}{\bigtriangleup}
\newcommand{\nn}{\nonumber}
\newcommand{\bed}{\begin{displaymath}}
\newcommand{\eed}{\end{displaymath}}
\newcommand{\bea}{\begin{eqnarray}}
\newcommand{\eea}[1]{\label{#1} \end{eqnarray}}
\newcommand{\beq}{\begin{eqnarray}}
\newcommand{\eeq}[1]{\label{#1} \end{eqnarray}}
\newcommand{\insertplotw}[1]{\centerline{\psfig{figure={#1},width=12.0cm}}}
\newcommand{\insertplot}[1]{\centerline{\psfig{figure={#1},height=4.0cm}}}


Our aim
in this work is a construction of an explicit dual model combining
direct channel resonances, Regge behaviour typical for hadrons and
scaling behaviour typical for the partonic picture. Such a model would produce 
a generalized $Q^2$-dependent dual amplitude $A(s,t,Q^2)$.
This amplitude, a function of three variables, should have correct
known limits, i.e. it should reduce to the on shell hadronic scattering
amplitude on mass shell, and to the nuclear structure
function (SF) when $t=0$. In such a way we could complete a unified
"two-dimensionally dual" picture of strong interaction
\cite{JM0dama,JM1dama,JMfitnospin,JMfitspin,MDAMA}:

\begin{figure}[htb]
\vspace*{-0.5cm}
        \insertplotw{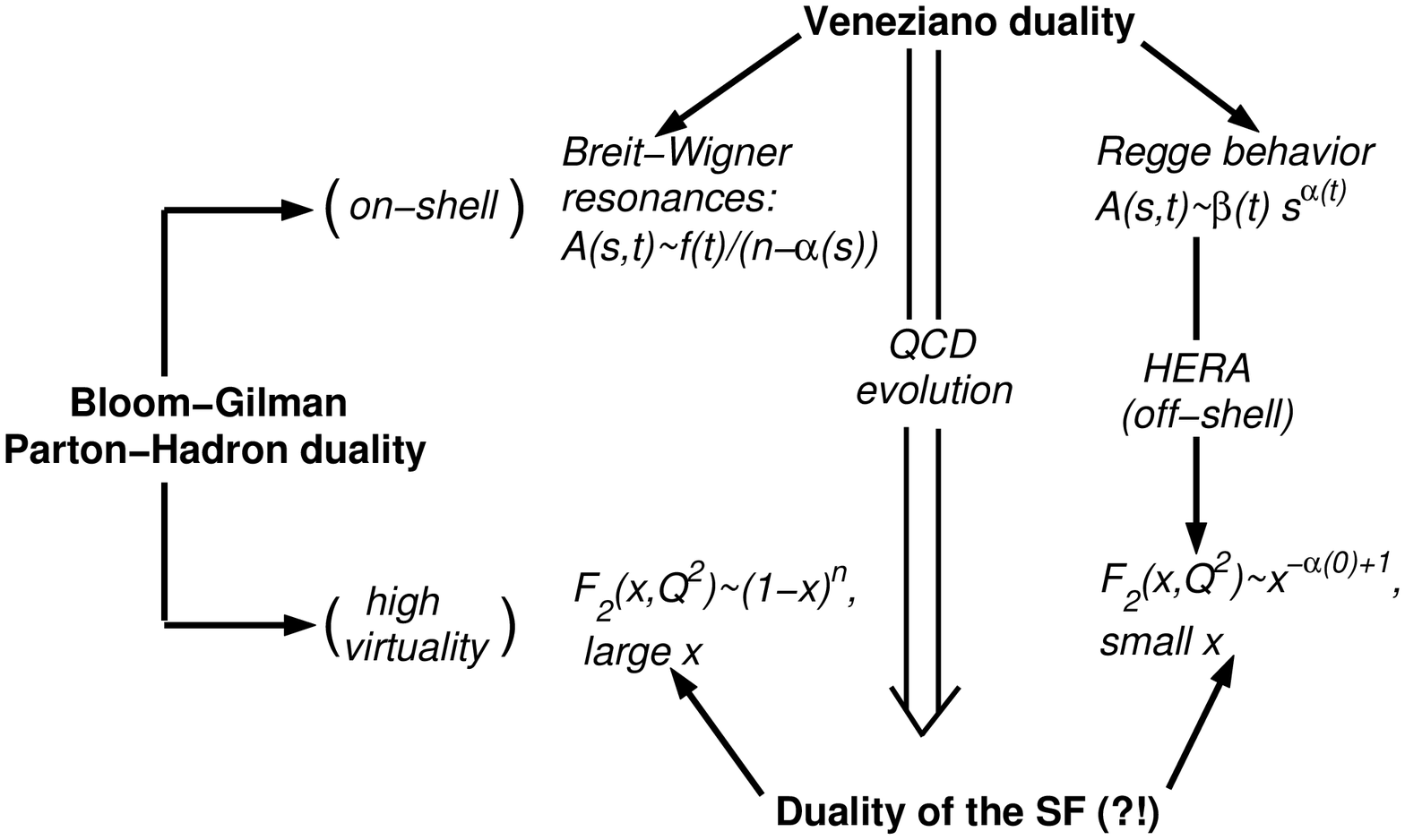}
\vspace*{-0.5cm}
\label{diag}
\end{figure}

About thirty years ago Bloom and Gilman \cite{BG}
observed that the prominent resonances in  inelastic
$e^-p$ scattering do not disappear with 
increasing photon virtuality $Q^2$, but fall at roughly the
same rate as background. Furthermore, the smooth scaling limit
proved to be an accurate average over resonance bumps seen at
lower $Q^2$ and $s$, this is so called Bloom-Gilman or hadron-parton duality. 
Since the discovery, the hadron-parton duality was studied in a number of papers 
\cite{carlson} and the new supporting data has come from the recent experiments 
\cite{Niculescu,Osipenko}.

First attempts to combine resonance (Regge) behaviour with
Bjorken scaling were made \cite{DG,BEG,EM} at low energies (large
$x$), with the emphasis on the right choice of the
$Q^2$-dependence, such as to satisfy the required behaviour of
form factors, vector meson dominance (the
validity (or failure) of the (generalized) vector meson dominance
is still disputable) with the requirement of
Bjorken scaling. Similar attempts in the high-energy (low $x$)
region became popular recently stimulated by the HERA data \cite{R8,BGP,K}.

A consistent treatment of the problem requires the account for the
spin dependence, which we ignore in this paper
for the sake of simplicity. Our goal is
rather to check qualitatively the proposed new way of constructing the
"two-dimensionally dual" amplitude.


In Refs. \cite{MDAMA} a modified
 definition of dual model with Mandelstam analyticity (M-DAMA)
with $Q^2$-dependence was proposed. M-DAMA preserves all the
attractive features of DAMA, such as pole decompositions in $s$
and $t$, Regge asymptotics etc., yet it gains the $Q^2$-dependent form factors, 
correct large and low $x$ behaviour for $t=0$ etc.
The M-DAMA integral reads \cite{MDAMA}:
\begin{equation}
D(s,t,Q^2)  =  \int_0^1 dz \biggl({z \over g}
\biggr)^{-\alpha_s(s')-\beta({Q^2}'')-1}
 \biggl({1-z \over
g}\biggr)^{-\alpha_t(t'')-\beta({Q^2}')-1}\,,
\label{mdama}
\end{equation}
where $a'=a(1-z)$, $a''=az$, and $g$ is a free parameter, $g>1$,
and $\alpha_s(s)$ and $\alpha_t(t)$ stand for the Regge
trajectories in the $s$- and $t$-channels. The $\beta(Q^2)$ is a smooth dimensionless function of $Q^2$,
which will be specified later on from studying different regimes
of the above integral.

The on mass shell limit,
$Q^2=0$, leads to the shift of the $s$- and $t$-channel
trajectories by a constant factor $\beta(0)$ (to be determined
later), which can be simply absorbed by the trajectories and, thus, 
M-DAMA reduces to DAMA. In the general case of the virtual particle
with mass $M$ we have to replace $Q^2$ by $(Q^2+M^2)$ in the
M-DAMA integral.

Now all the machinery developed for the DAMA model (see for example \cite{DAMA})
can be applied to the M-DAMA integral. Below we shall report briefly only some of
its properties, relevant for the further discussion.

The dual amplitude $D(s,t,Q^2)$ is defined by the integral
(\ref{mdama}) in the domain ${\cal R}e\
(\alpha_s(s')+\beta({Q^2}''))<0$  and ${\cal R}e\
(\alpha_t(t'')+\beta({Q^2}'))<0$. For monotonically decreasing
function ${\cal R}e\ \beta({Q^2})$ (or non-monotonic function with maximum at $Q^2=0$) and for 
increasing or constant
real parts of the trajectories these equations, applied
for $0\le z\le 1$, mean 
$
{\cal R}e\ (\alpha_s(s)+\beta(0))<0
$ 
and
${\cal R}e\
(\alpha_t(t)+\beta(0))<0 \,. $
To enable us to study the
properties of M-DAMA in the domains ${\cal R}e\
(\alpha_s(s')+\beta({Q^2}''))\ge 0$ and ${\cal R}e\
(\alpha_t(t'')+\beta({Q^2}'))\ge 0$, which are of the main interest, we
have to make an analytical continuation of M-DAMA. This leads to the appearance of 
two moving poles
\beq
\alpha_s(s(1-z_n))+\beta(Q^2z_n)=n\quad {\rm and} \quad 
\alpha_t(tz_m)+\beta(Q^2(1-z_m))=m, \quad n,m=0,1,2...
\eeq{poles}
The singularities of the dual amplitude are generated by pinches
which occur in the collisions of the above mentioned moving and fixed
singularities of the integrand $z=0,1$ \cite{DAMA}. 

It was shown in Refs. \cite{MDAMA}, that in this way for the $s$ poles (and similarly for $t$ poles)
we obtain the following expression for the pole term:
\begin{equation}
D_{s_n}(s,t,Q^2)=g^{n+1}\sum_{l=0}^{n}\frac{[\beta'(0)Q^2-s\alpha_s'(s)]^{l}C_{n-l}(t,Q^2)}
{[n-\alpha_s(s)-\beta(0)]^{l+1}}\,,
\label{p7}
\end{equation}
where
\begin{equation}
C_l(t,Q^2)=\frac{1}{l!}\frac{d^l}{dz^l}\left[\biggl({1-z \over
g}\biggr)^{-\alpha_t(tz)-\beta(Q^2(1-z))-1}\right]_{z=0}\,.
\label{p5}
\end{equation}
The modifications with respect to DAMA \cite{DAMA} are the shift of the 
trajectory $\alpha_s(s)$ by the factor of $\beta(0)$; and the coefficients $C_{l}$ are now 
$Q^2$-dependent and can be associated with the form factors. 

The new thing here is a possibility of having $Q^2$ poles, defined by
\begin{equation}
\alpha_s(0)+\beta(Q^2_n)=n\,, \quad  \alpha_t(0)+\beta(Q^2_m)=m\,.
\label{case2}
\end{equation}
In this
sense we can think about $\beta(Q^2)$ as of a kind of
trajectory. As
we will see later  with a proper choice of $\beta(Q^2)$ we can avoid 
unphysical poles in positive $Q^2$ region.


Below, for the first time,  the calculations for the $Q^2$-pole terms for the last case in eqs. (\ref{case2}) are presented. 

The point $z_m$ is a solution of the second 
equation in system (\ref{poles}) and for the pole condition, $z_m\rightarrow 0$,  it gives
\begin{equation}
\alpha_t(0)-t\,\alpha_t'(0)z_m+\beta(Q^2)-\beta'(Q^2)Q^2z_m=m\quad \Rightarrow \quad
z_m=\frac{m-\alpha_t(0)-\beta(Q^2)}{t\,\alpha_t'(0)-Q^2\beta'(Q^2)}\,.
\label{p12}
\end{equation}
Then residue at the pole $z_m$ is equal to:
\bea
2\pi i Res_{z_m} & =  & \frac{1}{t\,\alpha_t'(0)-Q^2\beta'(Q^2)}
\biggl({z_m \over g}
\biggr)^{-\alpha_s(s(1-z_m))-\beta(Q^2z_m)-1}
\biggl({1-z_m \over g}\biggr)^{-m-1} \nn \\
  & & =\frac{g^{m+2+\alpha_s(s)+\beta(0)}
[t\,\alpha_t'(0)-Q^2\beta'(Q^2)]^{\alpha_s(s)+\beta(0)}}
{[m-\alpha_t(0)-\beta(Q^2)]^{\alpha_s(s)+\beta(0)+1}}
F(z_m)\,,     
\eea{p13}
where the non-pole function is
\bea
F(z_m)&=&\biggl({z_m \over g} \biggr)^{-\alpha_s(s(1-z_m))+\alpha_s(s)
-\beta(Q^2z_m)+\beta(0)}
\cdot \frac{1}{(1-z_m)^{m+1}}\nn \\
& & \approx e^{\ln\left({z_m \over g} \right)  z_m\kappa}\frac{1}{(1-z_m)^{m+1}}\,,\quad 
z_m \ll 1\,, \quad \kappa=s\alpha'_s(s)-Q^2\beta'(0)\,.
\eea{p13a}
The $\kappa$ can be either positive or negative. 
Now as we have done above we would like to expand $F(z_m)$ in a series for 
$z_m \ll 1$, but there is a problem, since the derivatives 
$\left.\frac{d^n F(z_m)}{dz_m^n}\right|_{z_m=0}$ are divergent for 
any $n\ge 1$. On the other hand, we
can expand $F$ in a power series of $y=\ln\biggl({z_m \over g} \biggr)  z_m$ 
($y\rightarrow 0$, $z_m\rightarrow 0$). The inverse function $z_m=Z_m(y)$ is
very (infinitely) smooth, i.e. 
$\left.\frac{d^n Z_m(y)}{dy^n}\right|_{y=0}=0$ for $n=\overline{1,\infty}$. 
Thus, 
\begin{equation}
F(z_m)=\sum_{l=0}^{\infty}\frac{1}{l!}\left(\ln\left({z_m \over g} \right) 
z_m\kappa\right)^l\,.
\label{p14}  
\end{equation}
Inserting (\ref{p14}) into (\ref{p13})
we obtain the following expression for the pole term:
\bea
D_{Q^2_m}(s,t,Q^2)&=& g^{m+2+\alpha_s(s)+\beta(0)}
\sum_{l=0}^{M(s)+1}
\frac{[t\,\alpha_t'(0)-Q^2\beta'(Q^2)]^{\alpha_s(s)+\beta(0)-l}
[s\alpha'_s(s)-Q^2\beta'(0)]^{l}}{l!}\nn \\
& & \cdot \frac{\ln^{l} 
\left[\frac{m-\alpha_t(0)-\beta(Q^2)}{g(t\,\alpha_t'(0)-Q^2\beta'(Q^2))}\right]}
{[m-\alpha_t(0)-\beta(Q^2)]^{\alpha_s(s)+\beta(0)-l+1}}\,, 
\eea{p17}
where
\begin{equation}
M(s)=[\alpha_s(s)+\beta(0)]\,,
\label{p18}
\end{equation}
i.e. the integer part of $\alpha_s(s)+\beta(0)$. 
Comparing this expression with that for $s$-poles, eq. (\ref{p7}), 
we can say that the power of the $Q^2$-pole is $s$-dependent and not always integer and also distorted by the
logarithms. Contrary to eq. (\ref{p7}), 
the number of multipoles - $M(s)+2$ - is the same for all $m$. $M(s)$ is
growing with $s$, so the $Q^2$-poles become more
prominent for larger $s$.

The SF is related to the imaginary part of the scattering amplitude in the following way
\begin{equation}
F_2(x,Q^2)={4Q^2(1-x)^2\over{\alpha \left(s-m_N^2\right) (1+4m_N^2 x^2/{Q^2})^{3/2}}}\,{\cal I}m\  
A(s(x,Q^2),t=0,Q^2)\,,
\label{f2sigma}\end{equation}
where
$\alpha$ is the fine structure constant, $m_N$ is a nucleon mass.

In the low $x$ limit: $x\rightarrow 0$, $t=0$, $Q^2=const$, $s=Q^2/x\rightarrow\infty$, 
$u=-s$   M-DAMA gives the following expression  - see Refs. \cite{MDAMA} for details: 
\begin{equation}
{\cal I}m\  A(s,0,Q^2)|_{s\rightarrow\infty}\sim s^{\alpha_t(t)+\beta(0)+1}g^{\beta(Q^2)}\,.
\label{eq35a}
\end{equation}

According to the two-component duality picture \cite{FH}, both the
scattering amplitude $A$ and the structure function $F_2$ are the sums
of the diffractive and non-diffractive terms. At high energies
both terms are of the Regge type. For $\gamma^* p$ scattering only
the positive-signature exchanges are allowed. The dominant ones
are the Pomeron and  $f$ Reggeon, respectively. The relevant
scattering amplitude is as follows:
\begin{equation}
B(s,Q^2)=i\sum_k R_k(Q^2)\Bigl({s\over{m_N^2}}\Bigr)^{\alpha_k(0)},
\label{eq4}
\end{equation}
where $\alpha_k$ and $R_k$ are Regge trajectories and
residues and $k$ stands either for the Pomeron or for the Reggeon. 
In the phenomenological models which are used nowadays to fit $F_2$ data 
\cite{BGP,K,Niculescu,Osipenko,DS,JMfitnospin,JMfitspin} 
the
$Q^2$-dependence is introduced "by hands", via residues, 
which are fitted to the data. 
Now we have a model which contains $Q^2$-dependence from the very beginning and automatically gives
a correct behaviour of the residues.

Let us now come back to M-DAMA results. Using eqs.
(\ref{f2sigma},\ref{eq35a}) we get:
\begin{equation}
F_2\sim s^{\alpha_t(0)+\beta(0)} Q^2 g^{\beta(Q^2)}\,.
\label{n1}
\end{equation}
We propose $\beta(Q^2)$ in the following form \cite{MDAMA}
\begin{equation}
\beta(Q^2)=
-1-\frac{\alpha_t(0)}{\ln g} \ln \left(\frac{Q^2+Q_0^2}{Q_0^2}\right)\,.
\label{n3}
\end{equation}
where $Q_0^2$ is some  characteristic $Q^2$ scale. So finally we obtain
\begin{equation}
F_2(x,Q^2)\sim
x^{1-\alpha_t(0)}\Bigl({Q^2\over{Q^2+Q_0^2}}\Bigr)^{\alpha_t(0)}\,,
\label{eq41}
\end{equation}
where slowly varying factor
$\Bigl({Q^2\over{Q^2+Q_0^2}}\Bigr)^{\alpha_t(0)}$ is typical for
the Bjorken scaling violation (for example \cite{K}).

Now let us turn to the large $x$ limit. In this regime $x\rightarrow 1$,
 $s$ is
fixed, $Q^2=\frac{s-m^2}{1-x}\rightarrow \infty$ and
correspondingly $u=-2Q^2$. Following \cite{MDAMA} we obtain qualitatively correct behaviour
\begin{equation}
F_2 \sim (1-x)^2Q^4g^{2\beta(Q^2/2)} \sim
(1-x)^{2\alpha_t(0)\ln 2g/\ln g}\,.
\label{n5}
\end{equation}

Let us now study $F_2$ given by M-DAMA in the resonance region.
In the vicinity of the resonance $s=s_{Res}$ only the resonance
term $D_{Res}(s,0,Q^2)$ is important in the scattering amplitude and
correspondingly in the SF. 
Using  $\beta(Q^2)$ in the form (\ref{n3}) we obtain \cite{MDAMA}:
\begin{equation}
C_1(Q^2) =
\left(\frac{gQ_0^2}{Q^2+Q_0^2}\right)^{\alpha_t(0)}
 \left[\alpha_t(0)+\ln g \frac{Q^2}{Q^2+Q_0^2} - \frac{\alpha_t(0)}{\ln g}
\ln \left(\frac{Q^2+Q_0^2}{Q_0^2}\right)\right]\,.
\label{c1}
\end{equation}
The term $\left(\frac{Q_0^2}{Q^2+Q_0^2}\right)^{\alpha_t(0)}$ gives
the typical $Q^2$-dependence for the form factor (the rest is a slowly varying function of
$Q^2$).
If we calculate higher orders of $C_n$ for subleading resonances,
we will see that the $Q^2$-dependence is still defined by the same
factor $\left(\frac{Q_0^2}{Q^2+Q_0^2}\right)^{\alpha_t(0)}$. 

As we saw above the appearance and properties of 
$Q^2$ poles depend on
the particular choice of the function $\beta(Q^2)$, and for 
our choice, given by eq. (\ref{n3}), the unphysical $Q^2$ poles can be avoided. 

We have chosen $\beta(Q^2)$ to be a decreasing function, then, according to
conditions (\ref{case2}), there are no $Q^2$ poles in M-DAMA in
the domain $Q^2\ge 0$, if
\begin{equation}
{\cal R}e\ \beta(0)<-\alpha_s(0)\,, \quad {\cal R}e\
\beta(0)<-\alpha_t(0)\,.
\label{bt}
\end{equation}
We have already fixed $\beta(0)=-1$  and, thus, we
see that indeed we do not have $Q^2$ poles, except for the case of
supercritical Pomeron with the intercept $\alpha_P(0)>1$. Such a
supercritical Pomeron would generate one unphysical pole at $Q^2=Q^2_{pole}$
defined by equation
\begin{equation}
-1-{\alpha_P(0)\over \ln g}\ln
\left(\frac{Q^2+Q_0^2}{Q_0^2}\right)+\alpha_P(0)=0\quad
\Rightarrow \quad Q^2_{pole}=Q_0^2(g^{{\alpha_P(0)-1} \over
\alpha_P(0)}-1)\,. \label{Qp}
\end{equation}
Therefore we can conclude that M-DAMA does
not allow a supercritical trajectory - what is good property from the theoretical
point of view, since such a trajectory violates the Froissart-Martin
limit \cite{Frois}.

There are other phenomenological models which use dipole Pomeron with the
intercept $\alpha_P(0)=1$ and also fit the data \cite{R8,wjs88,Pom1}.
This is a very interesting case - ($\alpha_t(0)=1$) - for the
proposed model. At the first glance it seems that we should 
anyway have a pole at $Q^2=0$. It should 
result from the collision of
the moving pole $z=z_0$ with the branch point $z=0$, where 
$\alpha_t(0)+\beta(Q^2(1-z_0))=0$ in our case. Then, checking the conditions
for such a collision: 
$$
\alpha_t(0)-t\,\alpha_t'(0)z_0+\beta(Q^2)-\beta'(Q^2)Q^2z_0=0\ \Rightarrow \ 
z_0=\frac{-\alpha_t(0)-\beta(Q^2)}{t\,\alpha_t'(0)-Q^2\beta'(Q^2)}\,,
$$
we see
that for $t=0$ and for $\beta(Q^2)$ given by eq. (\ref{n3}) the
collision is simply impossible, because $z_0(Q^2)$ does not tend to $0$
for $Q^2\rightarrow 0$. Thus, for the Pomeron with
$\alpha_P(0)=1$ M-DAMA does not contain any unphysical
singularity.

On the other hand, a Pomeron trajectory with $\alpha_P(0)=1$ does
not produce rising SF, as required by the experiment.
So, we need a harder singularity and the simplest one is a dipole
Pomeron. A dipole Pomeron produces poles of the second power - 
$
D_{dipole}(s,t_{m})\propto
\frac{C(s)}{(m-\alpha_P(t)+1)^2}\,,
$ 
see for example Ref. \cite{wjs88} and
references therein. 

So far we have discussed DIS, where $Q^2$ is positive. However, in  $e^+ e^-$ annihilation we can access the negative $Q^2$ region, 
and according to vector meson dominance 
here we should have $Q^2$ poles corresponding to vector mesons. The behaviour of $\beta(Q^2)$ in this 
region is under investigation now. 

{\bf Conclusions:} \ \ A new model for the $Q^2$-dependent dual amplitude
with Mandelstam analyticity is proposed. The M-DAMA preserves all the attractive
properties of DAMA, such as its pole structure and Regge asymptotics, but it also
leads to generalized
dual amplitude $A(s,t,Q^2)$ and in this way realizes a
unified "two-dimensionally dual" picture of strong interaction
\cite{JM0dama,JM1dama,JMfitnospin,JMfitspin}.
This amplitude, when $t=0$, can be related
 to the nuclear SF, and in this way we fix the
 function $\beta(Q^2)$, which introduces the $Q^2$-dependence in M-DAMA, eq.
 (\ref{mdama}). Our analysis shows that for both large and low $x$
limits as well as for the resonance region the results of M-DAMA are in
qualitative agreement with the experiment. In this work as well as in \cite{MDAMA} 
only the positive $Q^2$ region was studied. The vector meson dominance suggests the possibility of having new features 
in the model - $Q^2$ poles in the negative $Q^2$ region, which is accessible in  $e^+ e^-$ annihilation. This study is in progress now.

{\bf Acknowledgments:} \ \ I thank  L.L. Jenkovszky and L. Lipatov for fruitful and enlightening discussions.

\end{document}